\documentclass[a4paper,11pt]{article}
\usepackage{jcappub} 
\usepackage{bm,graphicx,dcolumn,epstopdf,epsf, latexsym,mathbbol, dcolumn, amssymb,amsmath,color,slashed, mathrsfs,mathcomp, simplewick}
\usepackage[center]{subfigure}
\usepackage{multirow}
\usepackage{tabularx}
\usepackage{makecell}
\usepackage{color, soul}
\usepackage{hyperref}

\usepackage{lineno}

\arxivnumber{2305.12323} 
\title{\boldmath Constraints on the rotating self-dual black hole with quasi-periodic oscillations}

	\newcommand{\bq}{\begin{equation}}
	\newcommand{\eq}{\end{equation}}
	\newcommand{\bqn}{\begin{eqnarray}}
	\newcommand{\eqn}{\end{eqnarray}}
	\newcommand{\nb}{\nonumber}
	\newcommand{\lb}{\label}

	\newcommand{\CQG}{Class. Quantum Grav.}

\author{Cheng Liu${}^{a,b,d}$}

\author{Haiguang Xu${}^{a,b}$}

\author{Hoongwah Siew${}^{a,b}$}
 
\author{Tao Zhu${}^{c, d}$}

\author{Qiang Wu${}^{c, d}$}

\author{Yuanyuan Zhao${}^{a, b}$}

\affiliation{${}^{a}$School of Physics and Astronomy, Shanghai Jiao Tong University, 800 Dongchuan Road, Shanghai, 200240, China\\
	${}^{b}$Shanghai Frontiers Science Center for Gravitational Wave Detection, 800 Dongchuan Road, Shanghai 200240, China\\
	${}^{c}$Institute for Theoretical Physics \& Cosmology, Zhejiang University of Technology, Hangzhou, 310023, China\\
	${}^{d}$ United Center for Gravitational Wave Physics (UCGWP),  Zhejiang University of Technology, Hangzhou, 310023, China}

\emailAdd{hgxu@sjtu.edu.cn; Corresponding author}

\abstract{An impressive feature of loop quantum gravity (LQG) is that it can elegantly resolve both the big bang and black hole singularities. By using the Newman-Janis algorithm, a regular and effective rotating self-dual black hole(SDBH) metric could be constructed, which alters the Kerr geometry with a polymeric function $P$ from the quantum effects of LQG geometry. In this paper, we investigate its impact on the frequency characteristics of the X-ray quasi-periodic oscillations(QPOs) from 5 X-ray binaries and contrast it with the existing results of the orbital, periastron precession and nodal precession frequencies within the relativistic precession model. We apply a Monte Carlo Markov Chain (MCMC) simulation to examine the possible LQG effects on the X-ray QPOs. We found that the best constraint result for the rotating self-dual geometry from LQG came from the QPOs of X-ray binary GRO J1655-40, which establish an upper bound on the polymeric function $P$ less than $6.17\times 10^{-3}$ at 95\% confidence level. 
This bound leads to a restriction on the polymeric parameter $\delta$ of LQG to be 0.67.}

\begin{document}
\maketitle
\flushbottom

\section{Introduction}
\renewcommand{\theequation}{1.\arabic{equation}} \setcounter{equation}{0}

The prevailing theory of gravity, known as general relativity (GR), has had remarkable success in describing the laws of gravity in our universe, e.g., the prediction of gravitational waves \cite{LIGOScientific:2016aoc} and the Kerr spacetime for astrophysical black hole \cite{EventHorizonTelescope:2019dse, EventHorizonTelescope:2022wkp, Ghez:2008ms}. However, a few problems still remain unsolved, such as its incompatibility with quantum mechanics \cite{Matos:2022rvo, QG1, QG2}, as well as the existence of singularities where the known laws of physics do not apply, such as at the origin of the universe \cite{singularity1, singularity2} and inside black holes \cite{hawking}. To overcome these issues, various modified or quantum gravity theories have been proposed as possible replacements or extensions to GR. Therefore, whether these theories can replace GR depends on a series of experiments and observational tests.

Inspired by the remarkable achievements made in loop quantum cosmology \cite{Singh:2009mz,
Ashtekar:2011ni}, loop quantum gravity (LQG) provides an elegant resolution of the classical big bang and black hole singularities. A regular static spacetime metric, self-dual spacetime, has been obtained in the mini-superspace approach based on the polymerization procedure in LQG \cite{LQG_BH}. This spacetime has no curvature singularity and its regularity is determined by two parameters arising from LQG: the minimal area and the Barbero-Immirzi parameter. Furthermore, this spacetime exhibits a T-duality symmetry, which inspires its name as the self-dual spacetime \cite{Sahu:2015dea, Modesto:2009ve}. Thanks to increasingly in-depth and accurate observations of black holes, such as the highest resolution images of the M87* central black hole \cite{EventHorizonTelescope:2019dse} and the central black hole Sgr A* of the Milky Way \cite{EventHorizonTelescope:2022wkp}, and decades of monitoring of the orbit of the S2 star near the Sgr A* \cite{Ghez:2008ms}, black holes in LQG have attracted a lot of interest in recent years, see e.g., \cite{AOS18a,AOS18b,BBy18,ABP19,ADL20,Perez17,Rovelli18,BMM18,Ashtekar20, Gan:2020dkb} and references therein. 

One may naturally wonder if the LQG effects on the self-dual spacetime can produce any observable spatial or temporal features that can be verified or constrained by present or future experiments and observations. This question has motivated a multitude of studies in the past decades from various perspectives \cite{Alesci:2011wn, Chen:2011zzi, Dasgupta:2012nk, Hossenfelder:2012tc, Barrau:2014yka, Cruz:2015bcj, add1, add2, Cruz:2020emz, Santos:2021wsw, Liu:2020ola, Sahu:2015dea, Zhu:2020tcf, Virbhadra:2022iiy, Yan:2022fkr}. In particular, in \cite{Liu:2020ola}, the LQG effects on the shadow of a spinning black hole were examined in detail, and their observational consequences were compared with the latest Event Horizon Telescope (EHT) observation of the supermassive black hole M87*. The gravitational lensing in the self-dual spacetime has also been investigated, and the polymeric function from LQG has been constrained by the Geodetic Very-Long-Baseline Interferometry Data of the solar gravitational deflection of radio waves \cite{Sahu:2015dea}. Recently, the solar system test and S2 near Sgr A* orbital Celestial mechanics test of the self-dual spacetime have been carried out \cite{Zhu:2020tcf, Yan:2022fkr}, from which the observational bounds on the polymeric function $P$ of LQG are obtained as well. In addition to the above, phenomenological studies of other kinds of loop quantum black holes have also been extensively explored, see \cite{Liu:2021djf, Daghigh:2020fmw, Bouhmadi-Lopez:2020oia, Fu:2021fxn, Brahma:2020eos, Yan:2023vdg} and references therein.

Since the 1980s, the QPOs were discovered as a peculiar astronomical phenomenon \cite{1979Natur.278..434S}, high-precision X-ray timing observations of black hole X-ray binaries provide us with a new excellent approach to test the nature of gravity in strong gravitational fields and to gain a deeper understanding of the geometric structure of black hole spacetime \cite{Stella:1997tc, Stella:1998mq}. A typical X-ray binary system is composed of a donor star and a compact central object, either a black hole or a neutron star, which accretes material from its stellar companion, causing the accretion disk to reach X-ray emitting temperatures, so that we are able to observe the relativistic motion of matter in strong gravitational fields and probe the ultra-dense matter that makes up neutron stars. If QPOs occurs in these systems, which manifests the X-ray light from an astronomical object flickers about certain frequencies, the typical angular resolution of their imaging is sub-nano-arcseconds, exceeding the observation level of existing imaging instruments, but their observable rapid temporal X-ray variability allows us to indirectly map the accretion flow \cite{Ingram:2019mna, Remillard:2006fc}.

The QPOs of the observed transient light curve can be decomposed to multiple frequencies. One of the highly regarded models for the X-ray QPOs is the relativistic precession model \cite{Stella:1997tc, Stella:1998mq}, which posits that the closed orbits of test particles moving around a central compact object are characterized by three distinct frequencies: the orbital frequency, the radial epicyclic frequency, and the vertical epicyclic frequency, whose combinations are the observed QPOs frequencies. These accreting gas particles orbiting the central object at several or tens of gravitational radii emit X-ray signals carrying information about strong-field gravitational effects. The model was originally established to explain the high-frequency QPOs of neutron star X-ray binaries and was later extended to be used on stellar-mass black hole binaries \cite{Stella:1998mq}. 

Although observations of QPOs phenomena in black hole binaries are far less abundant than those in neutron stars, black holes provide a relatively clean astrophysical environment for studying the properties of gravity in strong fields and the geometry of black holes \cite{Motta:2013wga}. So far, several examples of observations of three model frequencies in black hole X-ray binaries using the relativistic precession model have been reported \cite{Motta:2013wga, Motta:2022rku, Ingram:2014ara}, while several sources with only two frequencies observed are informed \cite{Remillard:2006fc, Remillard:2002cy}. This field has been explored in many works from the theoretical point of view. For instance, testing the no-hair theorem with GRO J1655-40 \cite{Allahyari:2021bsq}, studying a black hole in non-linear electrodynamics \cite{Banerjee:2022chn} or the nature of the black hole candidate \cite{Bambi:2012pa, Bambi:2013fea}, investigating the QPO behaviors around rotating wormholes \cite{Deligianni:2021ecz, Deligianni:2021hwt}, testing gravity with the different kinds of modified gravity theories \cite{Maselli:2014fca, Chen:2021jgj, Wang:2021gtd, Jiang:2021ajk}, etc. Although there is still a lack of comparison between the observation of X-ray QPOs and the theory included the effects of LQG.

To address this issue, we investigate the effects of rotating SDBH spacetime from LQG on the QPOs of the X-ray binaries. We select five cases with well-measured observational data to explore and constrain the effects of LQG in rotating SDBH: GRO J1655-40 \cite {Motta:2013wga}, XTE J1550-564 \cite{Remillard:2002cy, Orosz:2011ki}, XTE J1859+226 \cite{Motta:2022rku}, GRS 1915+105 \cite{Remillard:2006fc} and H1743-322 \cite{Ingram:2014ara}. The effects of LQG may not only affect the X-ray signals emitted from the accretion disk around the black hole but also affect the orbital motion of the accreting material, especially the precession frequencies of the gas particle on the disk in the strong gravitational field spacetime of the black hole.  With the observational results of these stellar-mass black holes binaries, we use Monte Carlo simulation to explore the possible effects of LQG on the X-ray QPOs within the relativity precession model, find that the effects of LQG on the QPOs frequencies have a feeble contribution and give an upper limit value of the polymeric function $P$ that characterizes this effect at 95\% confidence level. 

Our paper is organized as follows. In Sec. II, we present a very brief introduction to the rotating SDBH. With this spacetime metric, in Sec. III, we present the derivation for the QPOs frequencies from the geodetic motion of a massive test particle in the rotating SDBH. In Sec. IV, we summarize the observation results from X-ray QPOs, describe the analysis of the MCMC method, and present our MCMC simulation best-fit results on constraining the parameters of the SDBH. We discuss the main results of our analysis. A summary and outlook of our works in this paper are presented in Sec. V.

	
\section{Brief introduction of the properties of rotating self-dual spacetime}\label{sec:LQBH}
 
In this section, we briefly introduce the rotating SDBH in LQG proposed in \cite{Liu:2020ola}. The self-dual spacetime arises from the symmetry-reduced model of LQG corresponding to homogeneous spacetimes. It has been shown to be geodesically complete and free of any spacetime curvature singularity.
The rotating SDBH solution in the Boyer-Lindquist coordinates is given by \cite{Liu:2020ola},
	\bqn
	d s^2 &=& \frac{\mathscr{H}}{\rho^2} \left[\frac{\Delta}{\rho^2}(dt-a\sin^2\theta d\phi)^2-\frac{\rho^2}{\Delta}dr^2-\rho^2d\theta^2 -\frac{\sin^2\theta}{\rho^2}(a dt-(k^2+a^2)d\phi)^2 \right],  \lb{mmm} \label{metric}
	\eqn
where the metric functions are defined as
	\begin{eqnarray}
	\Delta(r)&=& g(r)h(r)+a^2,\\
	k(r)&=& \sqrt{\frac{g(r)}{f(r)}}h(r),\\
	\rho^2(r)&=&k^2+a^2\cos^2\theta,
	\end{eqnarray}
	{where $a=J/M$ is the specific angular momentum (rotation parameter),  and $M$ and $J$ are the mass and angular momentum of the black hole, respectively}. It is easy to verify that when $a=0$ we recover the non-rotating SDBH solution \cite{Zhu:2020tcf}. The factor $\mathscr{H}$ in the metric \ref{metric} is a regular function which satisfied $h(r)=\lim\limits_{a \to 0} \mathscr{H}$, so the coefficient $\mathscr{H}/\rho^2$ is a normalization factor that does not  affect the intrinsic geometry of the spacetime. And here the metric auxiliary functions $f(r)$, $g(r)$ and $h(r)$ are given by
	\begin{eqnarray}
	f(r)&=&\frac{(r-r_+)(r-r_-)(r+r_*)^2}{r^4+a_0^2},\nonumber\\
	g(r)&=&\frac{(r-r_+)(r-r_-)r^4}{(r+r_*)^2(r^4+a_0^2)},\nonumber\\
	h(r)&=&r^2+\frac{a_0^2}{r^2}.
	\end{eqnarray}
where $r_+=2 G M/(1+P)^2$ and $r_{-} = 2G M P^2/(1+P)^2$ are the two horizons, and $r_{*}= \sqrt{r_+ r_-} = 2G MP/(1+P)^2$ with $G$ representing the gravitational constant, $M$ denoting the ADM mass of the SDBH, and $P$ being the polymeric function
\bqn
P \equiv \frac{\sqrt{1+\epsilon^2}-1}{\sqrt{1+\epsilon^2}+1}, \label{P_def}
\eqn
where $\epsilon$ is a product of the Immirzi parameter $\gamma$ and the polymeric parameter $\delta$, i.e., $\epsilon=\gamma \delta \ll 1$. The parameter $a_{0}$ is defined as
\bqn
a_0 = \frac{A_{\rm min}}{8\pi},
\eqn
where $A_{\rm min}$ denotes the minimum area gap of LQG. It is worth noting that $A_{\rm min}$ is related to the Planck length $l_{\rm Pl}$ through $A_{\min} \simeq 4 \pi \gamma \sqrt{3} l_{\rm Pl}^2$ \cite{Modesto:2009ve,Sahu:2015dea}.Therefore, $a_0$ is a quantity proportional to the Planck length $l_{\text{PI}}$ and could be ignored when considering an orbital motion on the scale of several to tens of Schwarzschild radii. For the parameter $\gamma$, different values are given in the literature from different considerations \cite{BenAchour:2014qca, Frodden:2012dq, Achour:2014eqa,
Han:2014xna,Carlip:2014bfa, Taveras:2008yf}. In this paper, in order to obtain the QPOs observational constraints of the polymeric function $\delta$ derived from the constraints on $P$, we adopt a value $\gamma=0.2375$ commonly used in the calculation of black hole thermodynamic quantities \cite{Meissner:2004ju}. To be able to reduce to the Newtonian limit, the gravitational constant in the rotating SDBH can be related to the Newtonian gravitational constant by 
\bqn
G_N=G\frac{(1-P)^2}{(1+P)^2}.
\eqn

Since the spin angular momentum of a SDBH also has a dimension of length, the gravitational constant in it needs to be similarly corrected. For convenience, in this paper, we take geometric units and set $G_N=1$. 

One can easily verify that when all the effects of LQG are absent ($P=0$), the metric of the rotating SDBH goes over to the Kerr spacetime exactly.

\section{QPOs frequencies in the rotating SDBH}

In this section, we derive the orbital, periastron precession, and nodal precession frequencies that describe QPOs within the relativistic precession model using the equations of motion of test particles on an accretion disk orbiting the rotating SDBH. The accretion disk is formed by particles moving in circular orbits around a compact object, whose physical properties and electromagnetic radiation characteristics are determined by the space-time geometry around the central compact object. For the purpose to study the fundamental frequencies that characterize the QPOs, let us first consider the evolution of a massive particle in the rotating SDBH spacetime. We start with the Lagrangian of the particle,
\bqn
\mathscr{L} = \frac{1}{2}g_{\mu \nu} \frac{d x^\mu} {d \lambda } \frac{d x^\nu}{d \lambda},
\eqn
where $\lambda$ denotes the affine parameter of the world line of the particle. For a massless particle, we have $\mathscr{L}=0$, and for a massive one $\mathscr{L} <0$. Then the generalized momentum $p_\mu$ of the particle can be obtained via
\bqn
p_{\mu} = \frac{\partial \mathscr{L}}{\partial \dot x^{\mu}} = g_{\mu\nu} \dot x^\nu,
\eqn
which leads to four equations of motions for a particle with energy $\tilde{E}$ and angular momentum $\tilde{L}$,
\bqn
p_t &=& g_{tt} \dot t +g_{t\phi} \dot \phi  = - \tilde{E},\\
p_\phi &=&g_{\phi t} \dot t+ g_{\phi \phi} \dot \phi = \tilde{L}, \\
p_r &=& g_{rr} \dot r,\\
p_\theta &=& g_{\theta \theta} \dot \theta.
\eqn
Here an overdot denotes the derivative with respect to the affine parameter $\lambda$ of the geodesics. From these expressions we obtain 
\bqn
\dot t =  \frac{g_{\phi\phi} \tilde{E}+g_{t\phi}\tilde{L}  }{ g_{t\phi}g_{\phi t}- g_{tt}g_{\phi\phi} } ~\\
\dot \phi = \frac{\tilde{E}g_{t\phi}+ g_{tt} \tilde{L}}{g_{tt}g_{\phi\phi}-g_{t\phi}g_{\phi t}}.
\eqn
For the conservation of the rest-mass, we have $ g_{\mu \nu} \dot x^\mu \dot x^\nu = -1$. Substituting $\dot t$ and $\dot \phi$ we can get
\bqn
g_{rr} \dot r^2 + g_{\theta \theta} \dot \theta^2 = -1 - g_{tt} \dot t^2  - g_{\phi\phi}\dot \phi^2 -2g_{t \phi}\dot{t}\dot{\phi}.~~
\eqn

Here we are interested in the evolution of the particle in the equatorial circular orbits. For this reason, we can consider $\theta=\pi/2$ and $\dot \theta=0$ for simplicity. Then the above expression can be simplified into the form
\bqn
\dot r ^2 = V_{\text{eff}}(r,M,\tilde{E},\tilde{L})  =\frac{\tilde{E}^2 g_{\phi\phi}+2\tilde{E}\tilde{L}g_{t\phi}+\tilde{L}^2 g_{tt}}{g^2_{t\phi}-g_{tt}g_{\phi\phi}} -1,              
\eqn
where $V_{\rm eff}(r)$ denotes the effective potential of the test particle with energy $\tilde{E}$ and axial component of the angular momentum $\tilde{L}$. The stable circular orbits in the equatorial plane are corresponding to those orbits with constant $r$, i.e., $\dot r^2=0$ and $dV_{\rm eff}(r)/dr=0$. With these conditions, one can write the specific energy $\tilde{E}$ and the specific angular momentum $\tilde{L}$ of the particle moving in a circular orbit in the black hole as 
\bqn
\tilde{E}&=&-\frac{g_{tt}+g_{t\phi}\Omega_\phi}{\sqrt{-g_{tt}-2g_{t\phi}\Omega_\phi-g_{\phi\phi}\Omega^2_\phi}} , \lb{Etilde}\\
\tilde{L}&=&\frac{g_{r\phi}+g_{\phi\phi}\Omega_\phi}{\sqrt{-g_{tt}-2g_{t\phi}\Omega_\phi-g_{\phi\phi}\Omega^2_\phi}},   \lb{ltilde}
\eqn
where the orbital angular velocity yields 
\bqn
\Omega_\phi=\frac{-\partial_r g_{t\phi}\pm \sqrt{(\partial_r g_{t\phi})^2-(\partial_r g_{tt})(\partial_r g_{\phi\phi})}}{\partial_r g_{\phi\phi}},~~~
\eqn
where the sign $\pm$ indicates corotating (counter-rotating) orbits. The corotating orbits have parallel angular momentum with spin while counter-rotating orbits have antiparallel ones.

On the other hand, since the test particles follow geodesic, equatorial, and circular orbits, we can also derive orbital angular momentum from the geodesic equations
\bqn
\frac{d}{d\lambda}(g_{\mu\nu}\dot x^\nu)-\frac{1}{2} (\partial_\mu g_{\nu\rho})\dot x^\nu \dot x^\rho=0.   \label{geo}
\eqn

With the conditions $\dot r=\dot \theta= \ddot{r}=0$ for equatorial circular orbits, the radial component of Eq.(\ref{geo}) reduces to 
\bqn
(\partial_r g_{tt})\dot t^2 +2(\partial_r g_{t\phi}) \dot t \dot \phi +(\partial_r g_{\phi\phi})\dot \phi^2=0.
\eqn
Therefore, we get the orbital angular velocity 
\bqn
\Omega_\phi=\frac{d\phi}{dt}=\frac{-\partial_r g_{t\phi}\pm \sqrt{(\partial_r g_{t\phi})^2-(\partial_r g_{tt})(\partial_r g_{\phi\phi})}}{\partial_r g_{\phi\phi}},~~~
\eqn
and the corresponding orbital frequency (or Keplerian frequency) is
\bqn
\nu_\phi = \frac{\Omega_\phi}{2\pi} =\frac{1}{2\pi}\Bigg(\frac{\sqrt{M}}{a_* M^{3/2}+r^{3/2}} 
+P\frac{\sqrt{M}}{r^{3/2}}+P\frac{M^{3/2}}{r^{5/2}}-6 a_* P\frac{M^2}{r^3}+\cdots \Bigg), \label{nphi}
\eqn
where $a_*\equiv a/M= J/M^2$.

Now let's consider a tiny perturbation around the circular equatorial orbit, we have
\bqn
r(t)=r_0 + \delta r(t), ~~ \theta(t)= \frac{\pi }{2} + \delta \theta(t),
\eqn
where the $\delta r(t)$ and $\delta \theta(t)$ are the tiny perturbations governed by the equations
\bqn 
\frac{d^2 \delta r(t)}{dt^2}+ \Omega_r^2 \delta r(t)=0, \\
\frac{d^2 \delta \theta(r)}{dt^2} +\Omega_\theta^2 \delta \theta(t) = 0,
\eqn 
where
\bqn
\Omega_r^2=-\frac{1}{2g_{rr}\dot{t}^2}\frac{\partial^2 V_{\text{eff}}}{\partial r^2}|_{\theta = \frac{\pi}{2}},\\ \Omega_\theta^2=-\frac{1}{2g_{\theta\theta}\dot{t}^2}\frac{\partial^2 V_{\text{eff}}}{\partial \theta^2}|_{\theta = \frac{\pi}{2}}.
\eqn

\begin{figure*} 
\centering
\includegraphics[width=5.0cm]{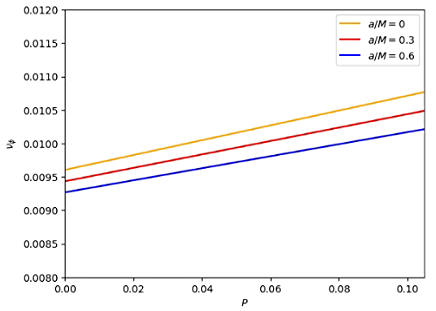}
\includegraphics[width=5.0cm]{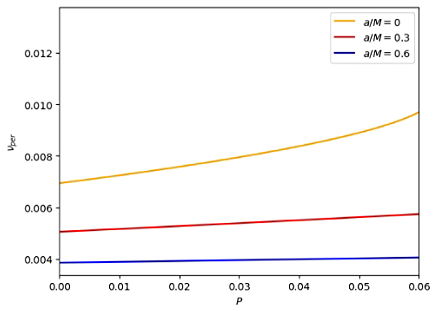}
\includegraphics[width=5.0cm]{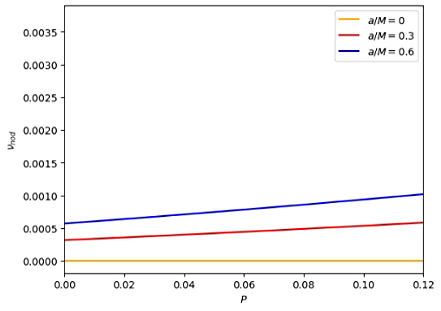}
\caption{The trend of the orbital frequency (or Keplerian frequency) $\nu_\phi$, periastron precession frequency $\nu_\text{per}$ and the nodal precession frequency $\nu_\text{nod}$ with respect to the polymeric function $P$ in rotating SDBH.}
\label{nu}
\end{figure*}
The radial epicyclic frequency $\nu_r$ and the vertical epicyclic frequency $\nu_\theta$ can be defined as $\nu_r=\Omega_r/2\pi$ and $\nu_\theta=\Omega_\theta/2\pi$, respectively. For the equatorial circular orbits of a test particle, the radial epicyclic frequency describes the radial oscillations around the mean orbit, and the vertical epicyclic frequency represents the vertical oscillations around the mean orbit. So the radial epicyclic frequency $\nu_r$ and the vertical epicyclic frequency $\nu_\theta$ in the rotating SDBH are given by
\bqn
\nu_r = \nu_\phi\Bigg[1-(6+8P)\frac{M}{r} + 8 a_* (1+5P) \frac{M^{3/2}}{r^{3/2}}-3a_*^2(1+8P)\frac{M^2}{r^2} \nb\\
+ 8a_* P\frac{M^{5/2}}{r^{5/2}} -8a_*^2 P\frac{M^3}{r^3}  +\cdots\Bigg]^{1/2}, \label{nr}
\eqn
\bqn
\nu_\theta=\nu_\phi\Bigg[1- 4a_*(1+5P)\frac{M^{3/2}}{r^{3/2}} + 3 a_*^2(1+8P)\frac{M^{2}}{r^{2}} -12a_* P\frac{M^{5/2}}{r^{5/2}}+ 16a_*^2 P \frac{M^3}{r^3}+\cdots \Bigg]^{1/2}.\label{ntheta}~~~~~
\eqn

One can immediately find out that, in the rotating SDBH spacetime, the frequencies $\nu_r$ and $\nu_\theta$ depend on the parameter $P$ from LQG. When the effect of LQG for the QPOs frequencies is not present ($P=0$), Eqs. (\ref{nphi}), (\ref{nr}), (\ref{ntheta}) naturally degenerates to the result of the Kerr black hole \cite{Ingram:2014ara}.

From the three fundamental frequencies discussed above, we can define the periastron precession frequency $\nu_{\text{per}}$ and nodal precession frequencies $\nu_{\text{nod}}$ as
\bqn
\nu_p=\nu_\phi - \nu_r, ~~~~~\nu_n = \nu_\phi -\nu_\theta.
\eqn
The variation of the orbital, periastron precession, and nodal precession frequency functions for the rotating SDBH with the parameter $P$ is shown in figure \ref{nu}. For an astrophysical black hole in general (which has a non-zero spin angular momentum), all three QPOs frequencies increase as the polymeric function increases. When the spin $a_*$ is small, the effect from the LQG contributes more significantly to the orbital frequency $\nu_\phi$ and the periastron precession frequency $\nu_r$, while the spin value $a_*$ is large, i.e. $a_*=0.6$, the contribution to the nodal precession frequency $\nu_{\text{nod}}$ is greater.

\section{Constraints on the SDBH parameter with current X-ray observations of QPOs}

In this section, we select 5 events of QPOs from different X-ray binaries with well-treated observational results, as presented in Table \ref{freq} to constrain the effects of LQG in rotating SDBH. We use the relativistic precession model along with the QPO frequencies from X-ray Observations of the GRO J1655-40, XTE J1550-564, XTE J1859+226, GRS 1915-105 and H1743-322 to make constraints on the parameters of the rotating SDBH. Finally, we report the best-fit results for traversing a reasonable physical parameter space using MCMC simulation methods.

\begin{table*} 
\renewcommand\arraystretch{1.5}
\caption{\label{binary}%
 The mass, orbital frequencies, periastron precession frequencies, and nodal precession frequencies of QPOs from the X-ray Binaries selected for analysis.}
\begin{tabular}{lccccc}
\hline\hline
  & GRO J1655-40 & XTE J1550-564 & XTE J1859+226 & GRS 1915+105 & H1743-322\\
  \hline
     $ M\; (M_{\odot})$ & 5.4$\pm$0.3 \cite {Motta:2013wga} & 9.1$\pm$ 0.61 \cite{Remillard:2002cy, Orosz:2011ki} & 7.85$\pm$0.46 \cite{Motta:2022rku} & $12.4^{+2.0}_{-1.8}$ 
 \cite{Remillard:2006fc}&  $\gtrsim 9.29$ \cite{Ingram:2014ara}\\
     $\nu_\phi$(Hz) & 441$\pm$ 2 \cite{Motta:2013wga} & 276$\pm$ 3 \cite{Remillard:2002cy} & $227.5^{+2.1}_{-2.4} $ \cite{Motta:2022rku} & 168 $\pm$ 3 
 \cite{Remillard:2006fc} &  
240 $\pm$ 3 \cite{Ingram:2014ara}\\
     $\nu_{\text{per}}$(Hz) & 298$\pm$ 4 \cite{Motta:2013wga} & 184$\pm$ 5 \cite{Remillard:2002cy} & $128.6^{+1.6}_{-1.8}$ \cite{Motta:2022rku} & 113 $\pm$ 5 \cite {Remillard:2006fc} & 
 $165_{-5}^{+9}$ \cite{Ingram:2014ara}\\
     $\nu_{\text{nod}}$(Hz) & 17.3$\pm$ 0.1 \cite{Motta:2013wga} & -  & $3.65\pm 0.01$ \cite{Motta:2022rku} & - & 
  $ 9.44\pm 0.02 $ \cite{Ingram:2014ara}\\ 
  \hline\hline
\end{tabular} \label{freq}
\end{table*}

\begin{figure*} 
\centering
\includegraphics[width=15.0cm]{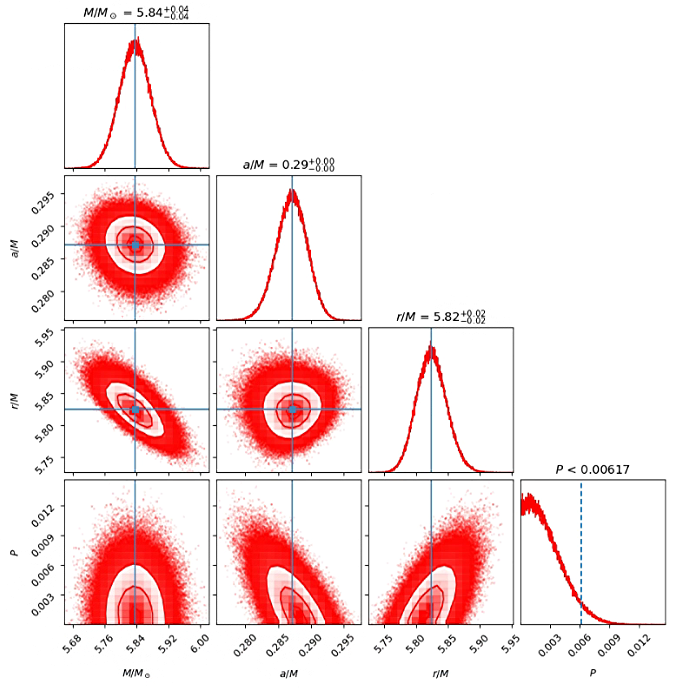}
\caption{Constraints on the parameters of the rotating SDBH with GRO J1655-40(red contours) from current observations of QPOs within the relativistic precession model.}
\label{contour3}
\end{figure*}

\begin{figure*} 
\centering
\includegraphics[width=7.2cm]{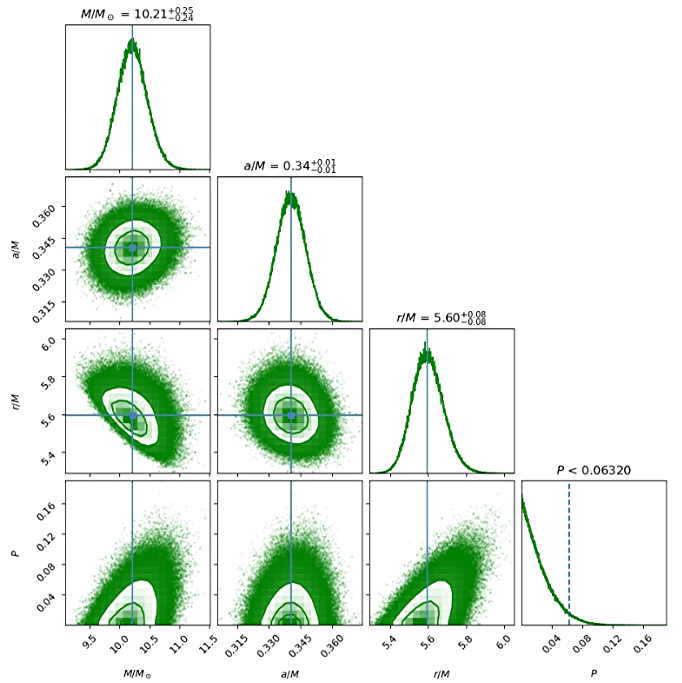}
\includegraphics[width=7.2cm]{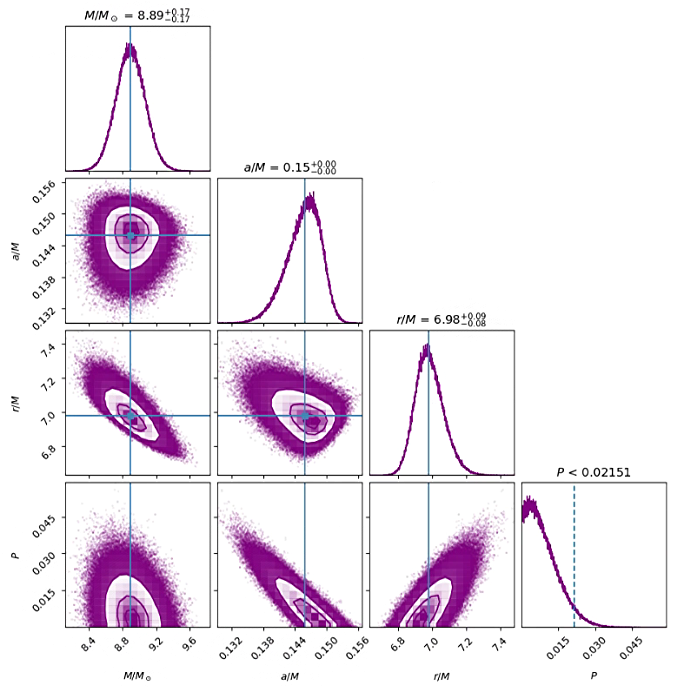}
\includegraphics[width=7.2cm]{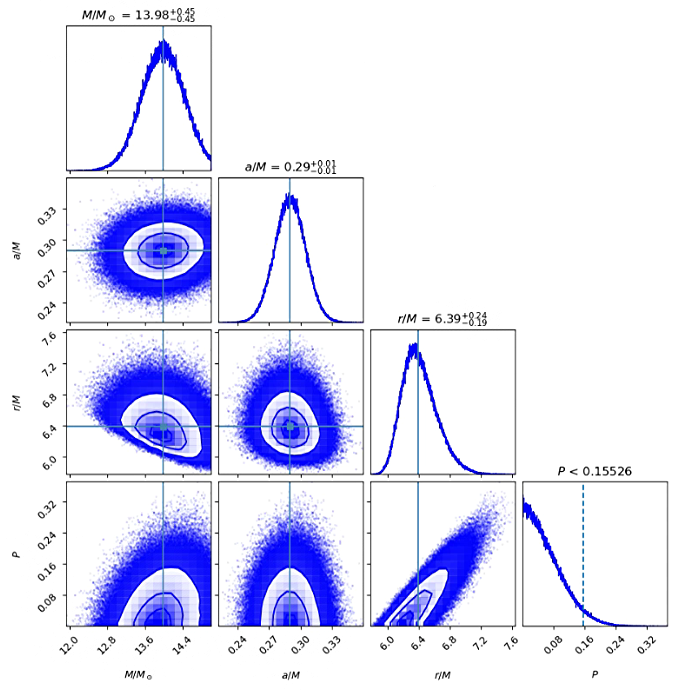}
\includegraphics[width=7.2cm]{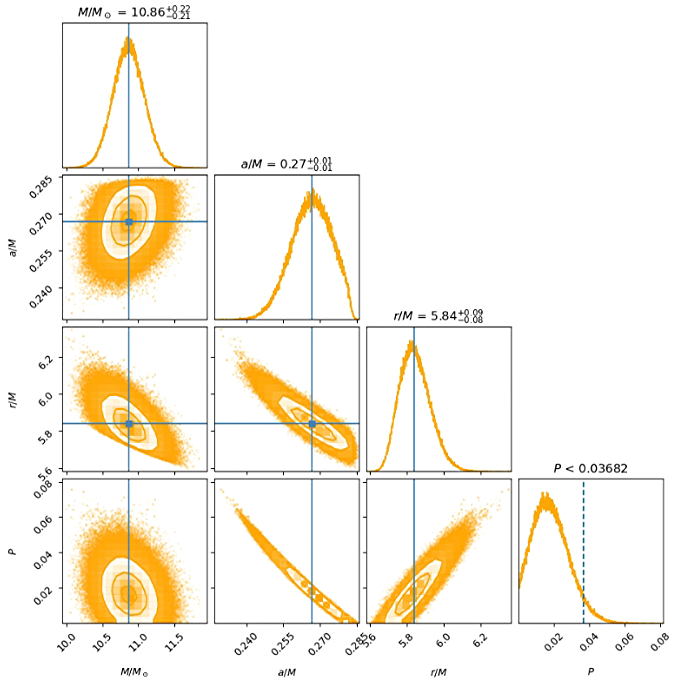}
\caption{Constraints on the parameters of the rotating SDBH with XTE J1550-564(green contours), and XTE J1859+226(purple contours), GRS 1915+105(blue contours), H1743-322(orange contours) from current observations of QPOs within the relativistic precession model.}
\label{contour}
\end{figure*}


\subsection{Analysis of Monte Carlo Markov chain}

\begin{table*}
\renewcommand\arraystretch{1.5} 
\caption{\label{prior}%
 The Gaussian prior of the rotating SDBH from QPOs for the X-ray Binaries.}
\begin{tabular}{lcccccccccc}
\hline\hline
\multirow{2}{*}{Parameters} & \multicolumn{2}{c}{GRO J1655-40}     & \multicolumn{2}{c}{XTE J1550-564} & \multicolumn{2}{c}{XTE J1859+226}    & \multicolumn{2}{c}{GRS 1915+105} & \multicolumn{2}{c}{H1743-322} \\
                            & $\mu$ & \multicolumn{1}{c}{$\sigma$} & $\mu$          & $\sigma$         & $\mu$ & \multicolumn{1}{c}{$\sigma$} & $\mu$         & $\sigma$         & $\mu$        & $\sigma$       \\
\hline
     $ M\; (M_{\odot})$ & $5.307$ & 0.066 & $9.10$ & 0.61 & $7.85$ & 0.46 & $12.41$ & 0.62 & $9.29$ & 0.46\\
     $a_*$ & $0.286$ & 0.003 & $0.34$ & 0.007 & $0.149$ & 0.005 & $0.29$ &   0.015 &$0.27$   & 0.013   \\
     $r/M$ & $5.677$ & 0.035 & $5.47$ & 0.12 & $6.85$ & 0.18 & $6.10$ & 0.30 & $5.55$ & 0.22 \\
     $P$ & \multicolumn{2}{c}{Uniform [0,1)}     & \multicolumn{2}{c}{Uniform [0,1)} & \multicolumn{2}{c}{Uniform [0,1)}    & \multicolumn{2}{c}{Uniform [0,1)} & \multicolumn{2}{c}{Uniform [0,1)} \\
     \hline\hline
\end{tabular}
\end{table*}

In this paper, we carry out the analysis of the MCMC implemented by \textit{emcee} \cite{emcee} to obtain the constraints on the rotating SDBH spacetime. The posterior can be defined as
\bqn
\mathcal{P}(\Theta|\mathcal{D},\mathcal{M})=\frac{P(\mathcal{D}|\Theta,\mathcal{M})\pi (\Theta|M)}{P(\mathcal{D}|\mathcal{M})}
\eqn
where $\pi(\Theta)$ is the prior and $P(D|\Theta,M)$ is the likelihood. The priors are set to be Gaussian prior within boundaries, i.e., 
$\pi(\theta_i) \sim \exp\left[{\frac{1}{2}\left(\frac{\theta_i - \theta_{0,i}}{\sigma_i}\right)^2}\right]$
, $\theta_{\text{low},i} < \theta_i < \theta_{\text{high},i}$, for paramaters $\theta_i = [M,a_*,r/M]$ and the $\sigma_i$ are their corresponding sigmas. We take the prior values of the parameters of the rotating SDBH as presented in Table \ref{prior}. For parameter $P$, we choose to use a uniform prior with a given boundary, i.e.  $\pi(P) = 1 \text{ for } P \in [P_\text{low}, P_\text{high}], \text{otherwise we set}$ $\pi(P) = 0$. According to the definition of the polymeric function (\ref{P_def}), setting $P\in [0,1)$ is a reasonable choice.

Following the orbital, periastron precession and nodal precession frequencies obtained in Sec.III, three different parts of data are employed in our MCMC analysis. So the likelihood function $\mathcal{L}$ can be written as
\begin{eqnarray}
\log {\cal L} = \log {\cal L}_{\rm obt} + \log {\cal L}_{\rm per} + \log {\cal L}_{\rm nod},\lb{likelyhood}
\end{eqnarray}
where $\log {\cal L}_{\rm obt}$ denotes the likelihood of the 3 astrometric orbital frequencies data
\bqn
 \log {\cal L}_{\rm obt} &=& - \frac{1}{2} \sum_{i} \frac{(\nu_{\phi\rm, obs}^i -\nu_{\phi\rm, th}^i)^2}{(\sigma^i_{\phi,{\rm obs}})^2} 
\eqn
and $\log {\cal L}_{\rm per}$ represents the likelihood of the data of the periastron precession frequency.
\begin{eqnarray}
\log {\cal L}_{\rm nod} =-\frac{1}{2} \sum_{i} \frac{(\nu_{\rm per, obs}^i -\nu_{\rm per, th}^i)^2}{(\sigma^i_{\rm per,{\rm obs}})^2},
\end{eqnarray}
and $\log {\cal L}_{\rm nod}$ is the likelihood of the nodal precession frequency
\bqn
\log {\cal L}_{\rm nod} =-\frac{1}{2} \sum_{i} \frac{(\nu_{\rm nod, obs}^i -\nu_{\rm nod, th}^i)^2}{(\sigma^i_{\rm nod,{\rm obs}})^2}.
\eqn
Here $\nu^i_{\phi,\rm obs}$, $\nu^i_{\rm per,\rm obs}$, $\nu^i_{\rm nod,\rm obs}$ are observation results of the orbital frequencies, periastron precession frequencies $\nu_{\text{per}}$ and nodal precession frequencies, and $\nu^i_{\phi,\rm th}$, $\nu^i_{\rm per,\rm th}$, $\nu^i_{\rm nod,\rm th}$ are the corresponding theoretical predictions, respectively. In the above expressions, $\sigma^i_{x, {\rm obs}^i}$ denote the corresponding statistical uncertainty for the associated quantities. 

We then ran the MCMC to constrain the value of parameters \{$M$, $a_*$, $r/M$, $P$\} for the rotating SDBH. In terms of the Gaussian prior, we rely on the parameter values given in the literature for the original data processing. For the QPOs cases that do not give explicit variances, e.g., GRS 1915+105, we take 5\% of the center value as our variance sigma, that is, assuming that its observation error has an accuracy of 5\%. We randomly sample $10^5$ points for each parameter according to the Gaussian prior distribution to traverse the physically possible parameter space within set boundaries to obtain the best fitting value of the parameters.

\subsection{Results and Discussions}

With the setup described in the above subsections, we explore the 4-dimensional parameter space for the rotating SDBH through an analysis of MCMC. The corresponding best-fit values of these 4 parameters are presented in Table.~\ref{binary}. Figures \ref{contour3}, and \ref{contour} show the MCMC analysis results of all parameters of the rotating SDBH for the 5 events of X-ray QPOs. On the contour plots of these figures, the shaded regions show the 68\%, 90\%, and 95\% confidence levels (C.L.) of the posterior probability density distributions of the entire set of parameters, respectively. 

As a result, we found that the best-fit values of $P$ come from the GRO J1655-40 (Fig.~\ref{contour}), which gives the upper limit of $P$ is 0.00617 at 95\% confidence level. 
The main reason is that, for the three fundamental frequencies required by the relativistic precession model, the data obtained from the observations of GRO J1655-40 are complete (all three frequencies observed), and their errors are smaller.  Similarly, complete observations of XTE J1859+226 and H1743-322 also yielded similar parameter constraint results.
The results presented in Fig.~\ref{contour3},  \ref{contour}, and Table.~\ref{binary} are fully consistent with a central black hole described by the Kerr spacetime as predicted by GR. We do not find any significant signature of the rotating SDBH spacetime.

We find the polymeric function $P$ can be constrained to be
\begin{eqnarray}
|P| < 0.00617,
\end{eqnarray}
at 95\% confidence level. This bound corresponds to the bound on the polymeric parameter $\delta$ of
\begin{eqnarray}
|\delta| < 0.67.
\end{eqnarray}

\begin{table*}
\renewcommand\arraystretch{1.5} 
\caption{\label{binary}%
 The best-fit values of the rotating SDBH parameters from QPOs for the X-ray Binaries.}
\begin{tabular}{lccccc}
\hline\hline
Parameters  & GRO J1655-40 & XTE J1550-564 & XTE J1859+226 & GRS 1915+105 &  H1743-322 \\
\hline
     $ M\; (M_{\odot})$ & $5.84^{+0.04}_{-0.04}$ & $10.21^{+0.25}_{-0.24}$ & $8.89^{+0.17}_{-0.17}$ &  $13.98^{+0.45}_{-0.45}$ & $10.86^{+0.22}_{-0.21}$\\
     $a_*$ & $0.29^{+0.00}_{-0.00}$ & $0.34^{+0.01}_{-0.01}$ & $0.15^{0.00}_{0.00}$ & $0.29^{+0.01}_{-0.01}$ &   $0.27^{+0.01}_{-0.01}$       \\
     $r/M$ & $5.82^{+0.02}_{-0.02}$ & $5.60^{+0.08}_{-0.08}$ & $6.98^{+0.09}_{-0.08}$ & $6.39^{+0.24}_{-0.19}$ & $5.84^{+0.09}_{-0.08}$ \\
     $P$ & $<0.00617$  &  $<0.06320$  &  $<0.02151$  &  $<0.15526$      &  $<0.03682$ \\
 \hline\hline
\end{tabular}
\end{table*}

Here we obtain a constraint that is slightly stronger (but still consistent) than the bound of $P<0.043$ derived in \cite{Yan:2022fkr} using the measurement data of the orbital procession of the S0-2 star, but much weaker than those obtained from the observation of the gravitational time delay by the Cassini mission, the deflection angle of light by the Sun, and the perihelion advance of Mercury \cite{Zhu:2020tcf}. Although the accuracy of the observational constraints from the QPOs is comparatively lower than that of other types of observations, our results demonstrate that the QPOs from X-ray binary observations have the capability of constraining the rotating black hole parameters beyond those supported by general relativity. Furthermore, we would like to mention that the tighter constraints discussed above are derived from observations at the scale of the solar system, while our results represent a bound on $P$ in a strong gravity regime environment with non-vanished angular momentum.

\section{Summary and Outlook}
\renewcommand{\theequation}{5.\arabic{equation}} \setcounter{equation}{0}

In this paper, we investigate the effects of the rotating SDBH from LQG on the QPOs. Firstly we obtain the three fundamental frequencies in rotating SDBH required by the relativistic precession model: orbital frequency, radial epicyclic frequency, and vertical epicyclic frequency. The influence of the LQG effect could not only change the precession shift of the accreting gas particles but also alter the fundamental QPOs frequencies of the X-ray binaries. Then we study the influence of the LQG effect with the observational results of the QPOs events from the X-ray binaries GRO J1655-40, XTE J1550-564, XTE J1859+226, GRS 1915+105 and H1743-322. With these observations, we perform an MCMC simulation to examine the potential LQG influence on the QPOs frequency, the periastron precession frequency, and the nodal precession frequency. We find that GRO J1655-40 give the best constraint results for their data is model complete and has a small error, and other results also offer an important reference for this constraint, especially the XTE J1859+226 and H1743-322, which are also model completely.  Among these results, we do not detect any significant indication of the rotating SDBH space-time  and thus establish an upper limit on the polymeric function $P\lesssim 0.00617$ at 95\% confidence level. This limit results in a constraint on the polymeric parameter $\delta$ of LQG to be $|\delta|\lesssim 0.67$. 

The sparsity of available observational data is the major limiting factor in constraining gravitational theories with QPOs.
In the future, we can take advantage of the multi-wavelength synergy of more advanced telescopes such as the planned Large Synoptic Survey Telescope (LSST) \cite{LSST:2008ijt}, Square Kilometre Array (SKA) \cite{2009IEEEP..97.1482D}, and BINGO (Baryon Acoustic Oscillations from Integrated Neutral Gas Observations) radio telescope \cite{Wuensche:2021dcx}. Their results have the potential to guide X-ray telescopes, i.e., Insight-HXMT(Hard X-ray Modulation Telescope) \cite{Lu:2019rru} and the next generation X-ray time-domain Telescope Einstein Probe \cite{2018SSPMA..48c9502Y}, to find more targeted observation sources and to take more accurate measurements. 
Furthermore, since X-ray binaries are also the origin of high-energy jets that are highly collimated \cite{Li:2022jxj}, fundamental frequencies of the X-ray QPOs and the relativistic precession of jets may be associated. Accurate measurements of jets' physical properties with SKA's superior sensitivity and angular resolution can therefore provide strong synergy for future QPOs observations.

\section*{Acknowledgements}

This work is supported by the Ministry of Science and Technology of China under Grant No. 2020SKA0110201 and the National Natural Science Foundation of China under Grant No. 11835009. T.Z. and Q.W. are supported by the Zhejiang Provincial Natural Science Foundation of China under Grants No. LR21A050001 and No. LY20A050002, the National Natural Science Foundation of China under Grants No. 11675143 and No. 11975203, the National Key Research and Development Program of China under Grant No. 2020YFC2201503, and the Fundamental Research Funds for the Provincial Universities of Zhejiang in China under Grant No. RF-A2019015.











\end{document}